\documentclass[11pt,twoside]{article}

\usepackage{asp2004}
\usepackage{epsf}
\usepackage{psfig}
\usepackage{lscape}

\newcommand{\msunyr}{\ensuremath{\mathit{M}_{\odot}{\rm yr}^{-1}}}   

\newcommand{\mdot}{\ensuremath{\dot{M}}}                             
\newcommand{\rstar}{\ensuremath{\mathit{R}_{\star}}}                 
\newcommand{\vinf}{\ensuremath{v_{\infty}}}                          

\usepackage{graphicx}
\usepackage[rightcaption]{sidecap}
\newcommand{\avep}{$\langle P \rangle$}

\begin{document}
\title{Using polarization to study the winds of massive stars}    
\author{Ben Davies$^{1,2}$, Jorick Vink$^3$, Ren\'{e} Oudmaijer$^1$}   
\affil{$^1$ University of Leeds, UK \\
  $^2$ R.I.T., USA \\
  $^3$ University of Keele, UK}    

\begin{abstract} 
The topic of wind-clumping has been the subject of much activity in
recent years, due to the impact that it can have on derived mass-loss
rates. Here we present an alternative method of investigating
wind-clumping, that of polarimetry. We present simulations of the
polarization produced by a clumpy wind, and argue that the
observations may be reproduced just by statistical deviations from
spherical symmetry when the outflow is only slightly fragmented. Here,
the polarization scales with $\dot{M}$, which is consistent with
observations of LBVs, WRs and O supergiants. Finally, we find clumping
factors in the inner 2$R_{\star}$ of $\sim 2-3$, and speculate as to
the clumping stratification of hot stars.
\end{abstract}

\section{Introduction}
Many massive stars show evidence of variable intrinsic polarization,
from OB supergiants \citep{L-N87}, to Luminous Blue Variables
\citep[LBVs, ][ \& refs therein]{Davies05}, and Wolf-Rayet stars
\citep[WRs, e.g. ][]{StL87}. This is typically attributed to
scattering off density-inhomogeneities, or `clumps', at the base of
the wind.

Clumping is an active research area at present due to the significant
effect it has on derived mass-loss rates when incorporated into model
stellar atmospheres (see e.g. Puls et al., this volume). Typically,
quantitative studies of clumping involve comparison of the synthetic
spectra produced by these model atmopsheres with observed data. Here
we present an alternative avenue of investigation into clumping, that
of polarimetry. Below, the basics of the technique are described. The
results of a recent spectropolarimetric survey of LBVs are reviewed,
and a quantitative investigation into the data is described.

\section{Studying clumping with polarization}
\begin{SCfigure}[][t]
  \includegraphics[width=6cm,bb=0 50 409 505]{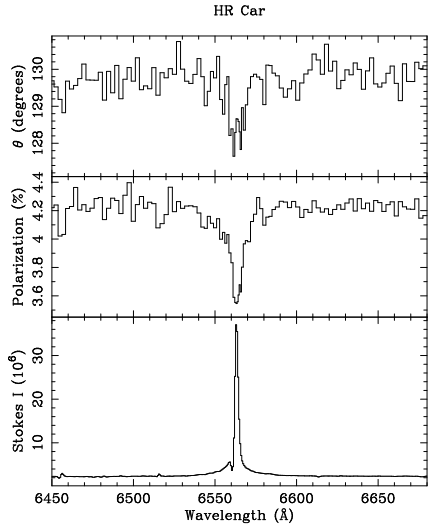}
  \caption{Polarization spectrum of the LBV HR Car. Bottom panel shows
  the intensity spectrum in the region of H$\alpha$, middle panel
  shows the degree of polarization, and top panel shows polarization
  position angle. The drop in polarization across the emission line is
  indicative of aspherical wind geometry on the plane of the sky.}
  \label{fig:examp}
\end{SCfigure}

Wind-asphericity in hot stars can produce intrinsic polarization via
the following mechanism: free electrons in the stellar wind scatter
continuum photons from the star, resulting in polarization of the
continuum perpendicular to the plane of scattering. If the overall
geometry of the scattering material is aspherical on the plane of the
sky, for example in a random distribution of clumps, this results in a
net continuum polarization. However, line photons which form over a
much larger volume undergo less scattering, meaning that the
line-emission remains essentially unpolarized. Therefore, by studying
polarization as a function of wavelength across strong emission lines,
wind-asphericity can be detected as a drop in polarization across the
emission line as the polarized flux is diluted. An example of this is
shown in Fig. \ref{fig:examp}.

\section{Spectropolarimetric survey of LBVs}
The above technique was applied to all known LBVs in the Galaxy and
Magellanic Clouds, originally with the aim of detecting disks /
bi-polar flows. It was found that at least half of those objects
studied showed evidence for wind-asphericity, and that this was more
frequent in the stars with the strongest H$\alpha$ emission. The 50\%
detection rate was deemed to be a lower-limit, due to the difficulty
in achieving the required S/N for the faintest objects. For more
details of this study, see \citet{Davies05}.

\begin{figure}[t]
\centering
  \includegraphics[height=5cm,bb=0 445 300 702,clip]{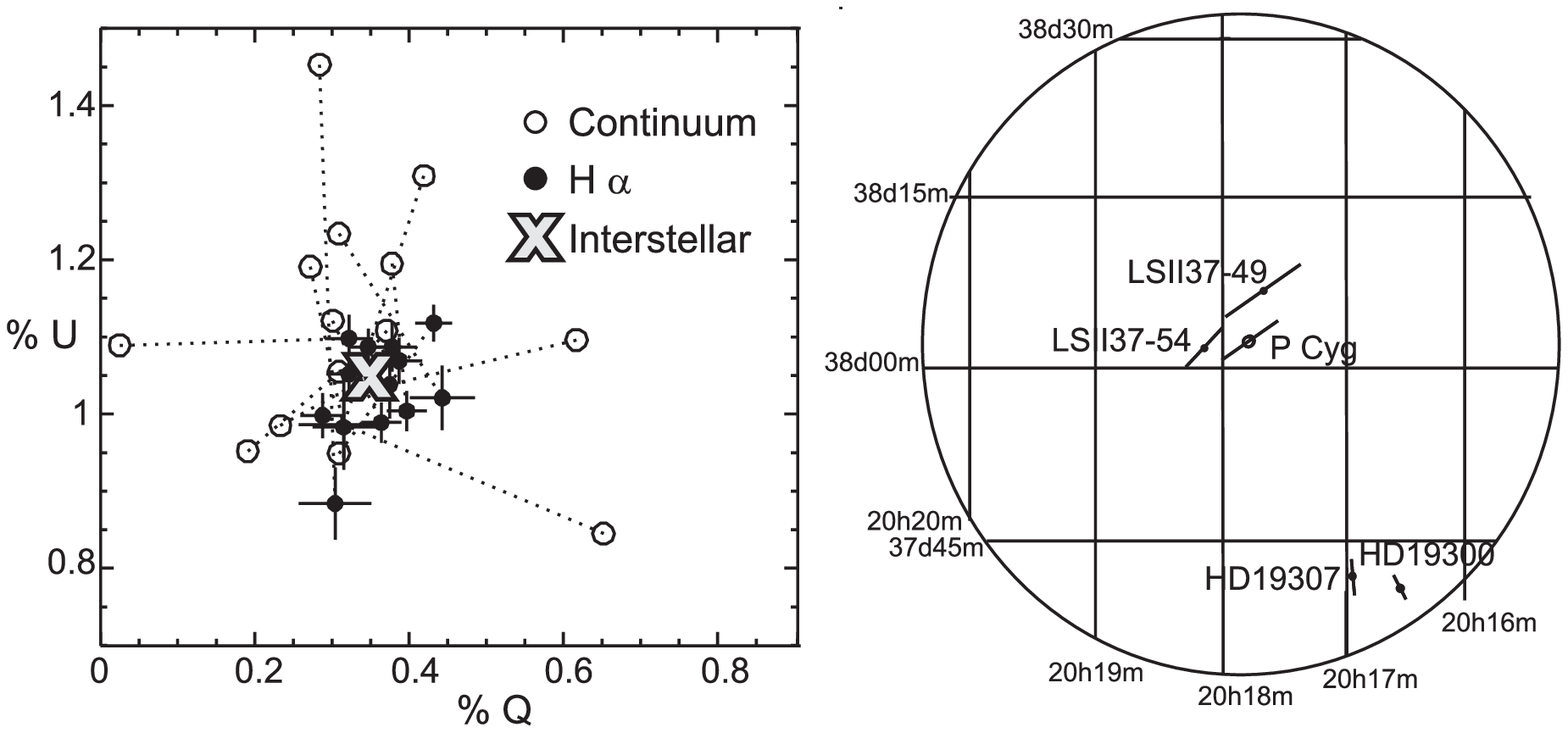}
  \includegraphics[height=5cm,bb=0 18 509 460,clip]{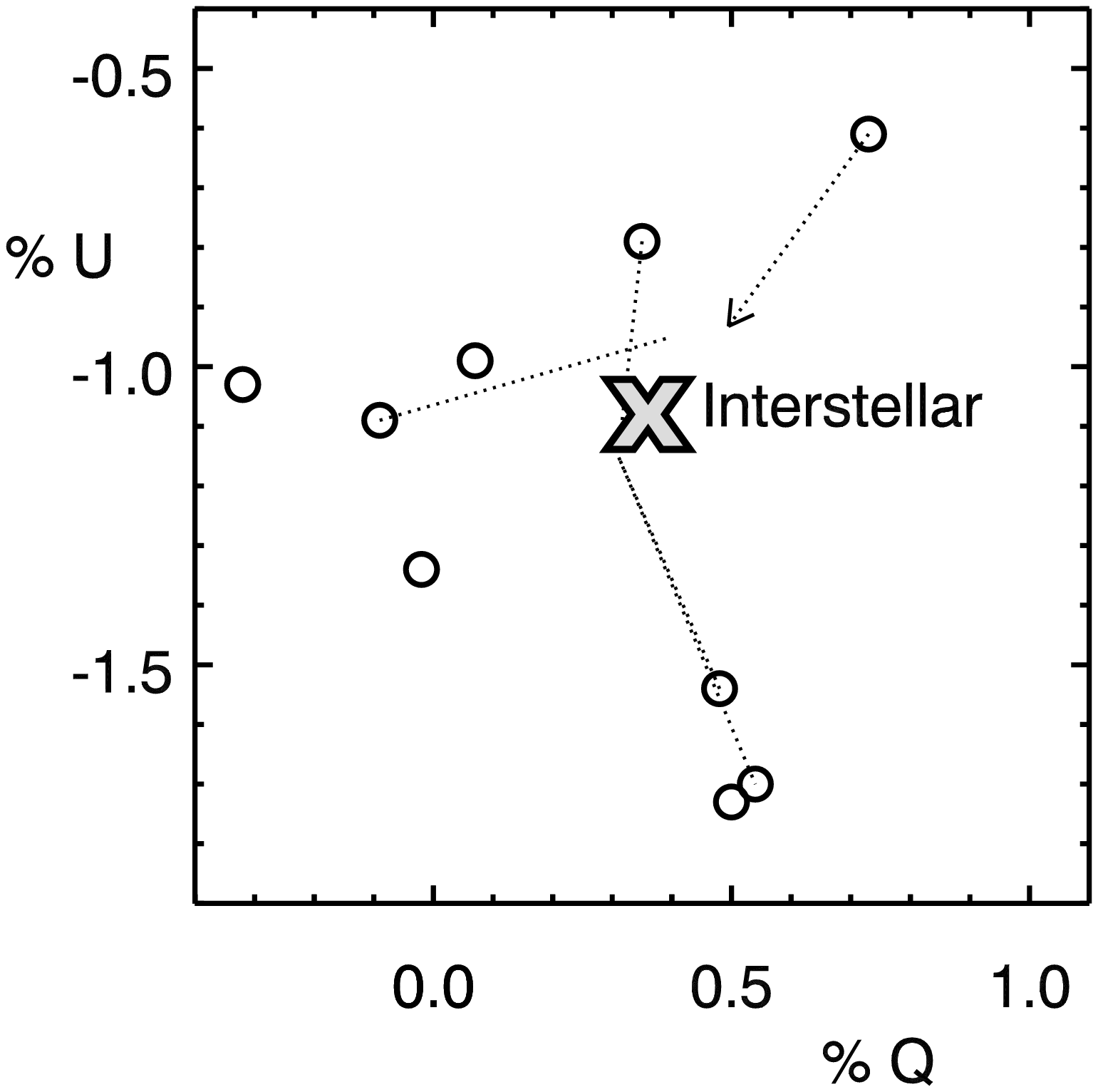}
\caption{Polarization vector diagrams of P Cyg \citep[{\it left},
][]{Nordsieck01} and AG Car \citep[][ {\it right}]{Davies05}. The
magnitude of the vectors between the continuum measurements (open
circles) and the interstellar polarization (i.e. the zero-point
polarization at the star) represent the strength of the intrinsic
polarization. The angle the vector makes with the $Q$ axis is a
function of PA. Dotted lines indicate the vector from the continuum
polarization to the that of H$\alpha$. It can be seen that the
polarization is variable in strength and PA, while the maximum
temporal resolution of the observations is of order days -- weeks.}
 \label{fig:qurand}
\end{figure}

It is clear that for those objects for which multiple observations
exist that the continuum polarization is variable, while the line
polarization remains roughly constant. This is strong evidence that
the H$\alpha$ emission is unpolarized, and that the continuum
polarization, which is variable in both strength and position-angle
(PA), is caused by electron-scattering within the line-forming
region. Examples of this behaviour are shown in Fig. \ref{fig:qurand}.

This behaviour is strong evidence of clumping at the base of the wind,
rather than wind axi-symmetry or binarity. The variability is
explained as follows: the polarization at any given epoch is due to
the total polarization of all clumps in the wind, and is dominated by
those clumps closest to the star. As these clumps move out through the
wind and new clumps are ejected, the polarization evolves. If repeat
observations are spaced such that the inner-wind bears no resemblance
to that at the previous observation, the polarization will have the
appearance of being completely random.

\section{Modelling the polarimetric variability}
To make a quantitative investigation of the results the LBV survey, as
well as polarimetry data of other hot stars, we have simulated the
polarimetric variability of a clumpy wind with a semi-analytic
model. Below is a brief outline of the model, a full description can
be found in Davies et al. ({\it in prep}).

Clumps are given a constant angular size, thickness and ejection
timescale. The clumps are assumed to move radially outwards according
to a standard velocity-law characterized by \vinf\ and acceleration
parameter $\beta$. By defining the parameter-space in this way, the
only free parameter for a given $R_{\star}$, $\dot{M}$ and
velocity-law was the ejection-rate per wind flow-time $\mathcal{N}$,
where the wind flow-time $t_{\rm fl} \equiv R_{\star} / \ v_{\infty}$.

For high $\mathcal{N}$, the clumps become less and less dense such
that the wind begins to approximate to a smooth outflow. For low
$\mathcal{N}$, the clumps have to become more dense in order to
conserve the mass-loss rate, and at some point become
optically-thick. At this point, the polarization-per-clump is assumed
to reach an asymptotic value, following the Monte-Carlo results of
\citet{R-M00}.

\begin{figure}[t]
\centering
  \includegraphics[width=8cm,clip]{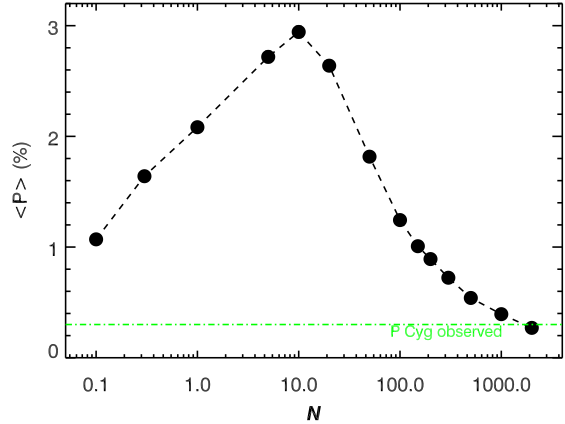}
  \caption{Time-averaged polarization of a clumpy wind as a function of
    clump ejection-rate per wind flow-time, using P Cyg stellar
    parameters. The dash-dotted line marks P Cyg's observed level of
    polarization. Taken from Davies et al. ({\it in prep}).}
    \label{fig:pcyg}
\end{figure}

Figure \ref{fig:pcyg} shows the time-averaged polarization \avep\ as a
function of $\mathcal{N}$ when the model is applied to P Cyg, using
the stellar/wind parameters derived by \citet{Najarro97}. The
predicted behaviour can be understood as follows: at high
ejection-rates, the wind consists of many low-density clumps which
tend to cancel each other out, producing low levels of
polarization. As the ejection-rate is decreased, the clumps must
become more and more dense, and the polarization-per-clump rises,
leading to higher overall levels of polarization. However, at some
point the clumps become optically-thick, and the maximum
polarization-per-clump is reached. Here, \avep\ begins to fall again as
the wind now consists of a fewer number of dense clumps.

It can be seen from Fig.\,\ref{fig:pcyg} that P Cyg's observed level
of polarization is consistent with two ejection-rate regimes:
$\mathcal{N} \la 0.1$ and $\mathcal{N} \ga 1000$. As the wind
flow-time of P Cyg is about 3 days, $\mathcal{N} = 0.1$ implies that
ejections occur only once per year. If ejections were really so
infrequent they would be evident in polarimetric monitoring
\citep[such as ][]{Hayes85}, and would be spectroscopically
conspicuous. Therefore, the high-$\mathcal{N}$ regime is preferred.

At ejection-rates of $\mathcal{N} \sim 1000$, the inner wind consists
of $\sim$7000 low-density clumps. One may expect that this situation
would result in no polarization, as the clumps cancel each other
out. However, only a slight imbalance is required to produce residual
polarization: the maximum polarization per clump here is 0.006\%, so a
Poissonian 1$\sigma$ deviation from spherical symmetry of around 80
clumps is enough to produce \avep$\sim$0.5\%. Therefore, the
polarization results from the statistical deviations from spherical
symmetry of a `fragmented' wind.

If intrinsic polarization arises naturally from a spherically
symmetric wind, then why don't all hot stars have it? The reason may
be due to the star's mass-loss rate -- for a given simulation with a
fixed ejection-rate, the time-averaged polarization is directly
proportional to the clump density, and hence the overall mass-loss
rate. If we were to reduce the input mass-loss rate of P Cyg from
$10^{-4.52}$\msunyr \citep[derived for P Cyg by ][]{Najarro97} to
$10^{-6}$\msunyr \citep[typical of O stars, ][]{Puls06}, \avep\ would
fall by a factor of 30 to $\sim$0.01\%. At this level the polarization
would be completely undetectable by current instruments.

The results of the modelling suggest two things:

\begin{itemize}
\item if the polarization is indeed produced by a wind that deviates
  only slightly from spherical symmetry and homogeneity, intrinsic
  polarization may be ubiquitous among hot stars with mass-loss rates
  greater than $\sim 10^{-4.5}$\msunyr. This is supported by the
  results of \citet{Davies05}, which showed that intrinsic
  polarization was more likely to be found among the LBVs with the
  strongest H$\alpha$ emission. It is also consistent with the results
  of \citet{Harries98,Harries02} who found a lower incidence of
  intrinsic polarization among O supergiants and WRs, which have
  $\dot{M} \sim 10^{-6} \rightarrow 10^{-5}$\msunyr.
\item the polarization must be variable on very short timescales.
\end{itemize}

In order to test this second conclusion, we have begun polarimetric
monitoring of the LBVs studied in \citet{Davies05}. By switching to
broad-band polarimetry, the throughput can be improved such that many
objects can be studied per night, while the precision of $\sigma P
\sim$0.03\% can be reached.

\begin{figure}[t]
\centering
  \includegraphics[width=8cm,bb=0 250 500 700,clip]{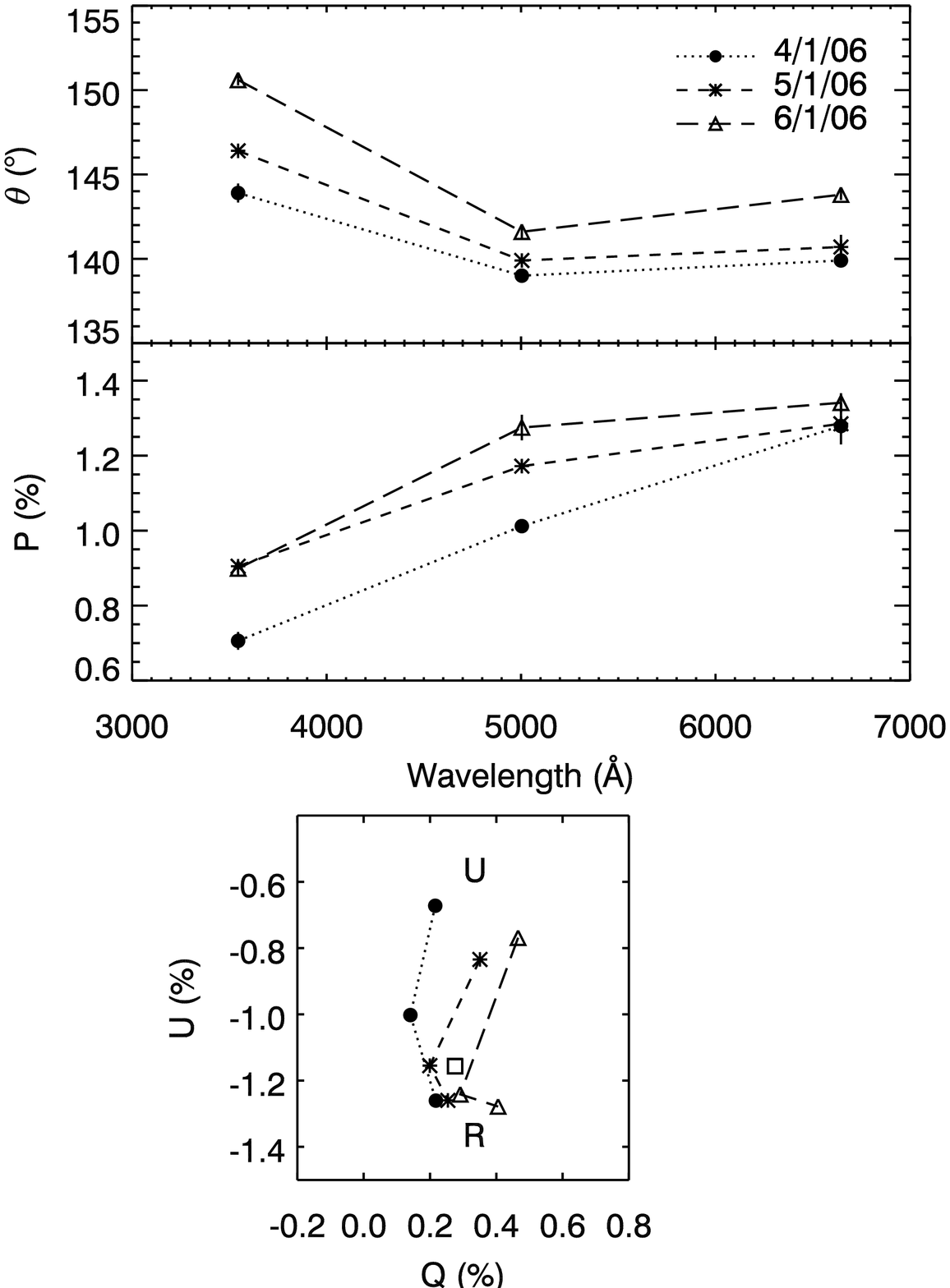}
\caption{Broad-band polarimetry of AG Car over three consecutive
  nights. A jump in polarization between the first and second nights
  ({\it unpublished}). }
 \label{fig:agcar}
\end{figure}

Figure \ref{fig:agcar} shows the broad-band polarimetric variability
of AG Car over three nights. It can be seen that the polarization
jumps by $\sim$0.2\% between the first and second night, particularly
in the $U$ and $V$ bands. As these observations were part of a short
pilot study, it is not clear if nightly variations of 0.2\% are
typical, as one would expect from the thousands of clumps in a
`fragmented' wind, or are rare events associated with the ejection of
dense clumps. We are presently obtaining polarimetric monitoring data
over longer time baselines to investigate this.

\section{Clump `filling-factor'}
In order to quantify wind-clumping in model atmospheres, the {\it
filling-factor} is defined as $f = \langle \rho^{2} \rangle / \langle
\rho \rangle ^{2}$. Under this formalism, the density of a clumped
wind, and therefore the mass-loss rate, is overestimated by a factor
$\sqrt{f}$. Traditionally, the behaviour of $f$ with distance from the
star $r$ is assumed to be that of increasing from unity at
$r=R_{\star}$ to some asymptotic value $f_{\infty}$
\citep[e.g. ][]{Dessart00}. It was found by \citet{Hillier03} and
\citet{Bouret05} that the clumping must reach $f_{\infty}$ very close
to the sonic point of the wind; and \citet[][ these
proceedings]{Puls06} have shown recently from combined H$\alpha$, IR
and radio data that clumping may be strongest in the inner
$\sim$15$R_{\star}$, tending towards a smoother outflow at larger
radii.

From our models, we have attempted to make an independent measurement
of the clumping of the inner wind by calculating the quantity $\langle
\rho^{2} \rangle / \langle \rho \rangle ^{2}$ for a given
simulation. It is assumed that there is no inter-clump medium, and the
clumps have mass and angular size which are constant with
distance. This is unrealistic, as hydrodynamical models of
radiatively-driven winds predict a wind structure which evolves with
distance from the star \citep{R-O02}. However, these assumptions are
valid in terms of this model as the polarization is insensitive to
material greater than $\sim 5 R_{\star}$. We therefore restrict our
analysis to the clumping in the inner wind region $r_{\rm in} = 1.05
\rightarrow 2$ \citep[as defined in][]{Puls06}.

\begin{figure}[t]
  \centering 
  \includegraphics[width=8cm,clip]{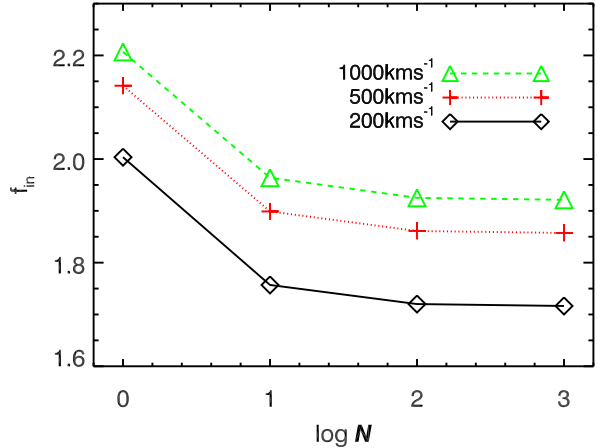}
  \caption{Clump filling-factor of the inner wind $f_{\rm in}$ as
    defined in \citet{Puls06} as a function of ejection rate for typical
    LBV terminal wind velocities. }
  \label{fig:f}
\end{figure}

Figure \ref{fig:f} shows $f_{\rm in}$ as a function of ejection rate
per wind flow-time $\mathcal{N}$ for typical LBV wind velocities and a
$\beta = 1$ velocity law. It can be seen that $f_{\rm in}$ is
increased for lower ejection rates and higher terminal velocities, due to
the material in the wind becoming more spread-out. 

If extrapolated to higher \vinf\ and lower \rstar\, the results of
$f_{\rm in}$ from this model (2-3) are comparable to those derived by
\citet{Puls06} for O stars (3-6, under the assumption that the outer
wind is unclumped). As the model presented here predicts a small
dependence on \vinf, an extension of the study by Puls et al. to
cooler supergiants would be a strong test of the model's validity.

\subsection{Clumping stratification}
\citet{Brown95} showed that the radial redistribution of material
above the photosphere (by e.g. radiative instabilities) could {\it
not} produce polarization. This can be understood as follows: if we
take a spherically-symmetric shell and compress it in the radial
direction, the optical depth along the thickness of the shell is
unchanged. Hence, if a region of a smooth outflow is radially
compressed, it will scatter no more light than the rest of the
outflow. Therefore, the polarimetric variability of not just LBVs, but
also WRs and O stars, is strong evidence that clumping begins at or
below the photosphere, and is not due to radiative instability-induced
clumping.

We find that, for typical LBV parameters, the clumping in the inner
wind (\rstar\ $\la$ 2) is in the region of $\sim$2. Our model makes no
attempt to calculate the dynamical evolution of the clumps, and hence
the evolution of $f$ with distance to the star. However, at larger
radii where the Sobolev length is large, radiative instabilities could
further coagulate an already clumped wind, leading to an increase in
$f$ with radius. This is consistent with the results of Puls et
al. (2006), who find an increase in $f$ beyond \rstar=2, at least for
stars with dense winds.

\section{Summary \& conclusions}
We show that, through observational monitoring and quantitative
modelling, polarimetry can be used to independently investigate the
phenomenon of wind-clumping in hot stars. From modelling the
polarimetric variability, it is found that the observations are
reproduced when the wind consists of a large number of low-density
clumps. Here the polarization can arise from statistical deviations
from spherical symmetry in only a slightly `fragmented' wind. 

It is predicted that polarization scales linearly with mass-loss rate,
and so is consistent with result of higher polarization in LBVs than
WRs and O supergiants. Short-timescale variability is also predicted,
and this has been detected in a polarimetric-monitoring
pilot-study. Further observations are planned. 

In an investigation of the wind clumping-factor $f_{cl}$, we find that
the wind is already significantly clumped in the inner 2$R_{\star}$,
in agreement with recent combined H$\alpha$/IR/radio observations. The
model could be further tested by extending these observations to B/A
supergiants. 

\acknowledgements 
We would like to thank Qingkang Li, Rico Ignace and John Brown for
many extremely useful discussions throughout the course of this work.



\end{document}